\documentstyle[12pt]{article} %available in the standard LaTex package.
\normalbaselines

\begin{document}
\bibliographystyle{unsrt}

\newcommand{\be}{\begin{equation}}
\newcommand{\ee}{\end{equation}}

\begin{center}
{\Large {\bf Experiment on interaction-free measurement in neutron
interferometry}}

\vspace{4mm}

Meinrad Hafner$^a$\footnote{Email: meinrad@galileo.ast.univie.ac.at}
and Johann Summhammer$^b$\footnote{Email: summhammer@ati.ac.at}

\vspace{4mm}

$^a$ Institute for Astronomy

University of Vienna

Tuerkenschanzstrasse 17

A-1180 Vienna, Austria

\vspace{2mm}

$^b$ Atominstitut der \"Osterreichischen Universit\"aten

Sch\"uttelstrasse 115

A-1020 Vienna, Austria

\end{center}

\begin{abstract}
A neutron interferometric test of interaction-free detection of the
presence of an absorbing object in one arm of a neutron interferometer
has been performed. Despite deviations from the ideal performance
characteristics of a Mach-Zehnder interferometer it could be shown that
information is obtained without interaction.
\end{abstract}

\section{Introduction}

In classical mechanics the interaction between a measuring device and
the object is treated as arbitrarily small and can therefore be
neglected. This is fundamentally different in quantum physics, as
examplified in the Heisenberg-microscope: A measurement of the position
of an electron inevitably leads to a disturbance of its momentum by the
scattering of a photon. It seems that no measurement can be performed
without the interaction with a particle.
Even in the case investigated by Dicke \cite{Dicke}, where knowledge
about the position of an atom is gained by ascertaining where it is not,
it had to be concluded, that photon scattering still is responsible for
providing the information. However, the scattering remains unresolvable,
because the change in the photon's momentum is within the momentum
uncertainty of the illuminating beam of light. But recently Elitzur and
Vaidman (EV) proposed an experiment where the presence of a perfectly
absorbing classical object  in one arm of a Mach-Zehnder interferometer
can be detected \cite{EV1}, \cite{EV2}, \cite{Bennet},
\cite{Vaidman.quant-ph}. EV symbolized the object by a bomb which
explodes when the measuring particle hits it. Their scheme is shown in
Fig.1. Here, no interaction is involved. For, had there been an
interaction, the test particle would have been absorbed and could not
have reached any of the detectors at the outputs of the interferometer.
The drawback of this scheme is that even when adjusting the
transmissivity of the beam splitters of the interferometer, the
probability of recognizing the object without interaction does not
exceed 50\%. This disadvantage could be overcome by Zeilinger et al.,
who introduced a multi-loop interferometer. In the limit of infinitely
many loops, one can obtain a probability of 100\% of determining the
presence of the absorbing object without interaction
\cite{Zeilinger.idea}. Paul and Pavicic incorporated this idea into a
proposal employing an optical resonator \cite{Paul.Pavicic}. A first
experiment with polarized photons has already been reported
\cite{Zeilinger.exp}, \cite{Kwiat.review}. Due to imperfections of
the optical devices only around 50\% of interaction-free detections
of the object were possible. The theoretical analysis of these
situations has also been extended to absorbing {\it quantum} objects,
where it is no longer possible to assume a perfect absorber. But even
then, a sizable amount of knowledge is gained without interaction
\cite{Krenn}, \cite{Karlsson}.

In this paper we report on the first experiment with neutrons. The
original scheme of EV was implemented in a single crystal neutron
interferometer. Although such interferometers are not ideal
Mach-Zehnder interferometers, it was possible to ascertain the presence
of an
absorbing object, which was realized as a neutron detector, with higher
probability than classical physics would permit, so that one must
conclude that some knowledge is obtained without interaction. Before
dealing with the details of this experiment, it is useful to recall the
ideal case of EV (Fig.1).

The two interferometric paths have equal length and the beam splitters
at the entrance and at the exit have the same reflectivity of 50\%.
When the interferometer is empty the probability amplitudes leading to
P1 interfere constructively, while those leading to P2 interfere
destructively. Therefore a particle will always be directed towards P1
(Fig.1a). If a perfectly absorbing object --- from now on called a black
object --- is inserted into one path of the
interferometer, only one beam can reach the exit ports, because the
wavefunction behind the object vanishes. No interference can occur.
With semitransparent mirrors at the entrance and exit of the
interferometer, there is a probability of 50\% that the particle will
be absorbed in the object, and a probability of 25\% each that it will
reach either P1 or P2.

In order to clarify why interaction-free measurement is involved here,
consider a collection of identically looking objects, some perfectly
absorbing, some perfectly transparent (i.e. not even inducing a phase
shift) for the kind of particles used in the interferometer. Will it be
possible to separate the objects into the two classes when testing them
in the interferometer? Classically there is no way. Without permitting
interaction we could only randomly pick out objects and put them into
the one or the other class. But quantum mechanically a separation is
possible. This is illustrated in Fig.2. Suppose we place one such object, of %%@
which it is not known
whether it is black or transparent, into one path of the
interferometer and send one particle. Three different outcomes can
occur:

\begin{description}
\item[i.] The particle is detected at port P1. This can happen in either
case, so no information is gained and another particle may be used.
\item[ii.] The particle is detected at port P2. This is impossible with
a perfectly transparent object, so we know now, that a black object is
in the interferometer. Only one particle entered the
interferometer and it was not absorbed, but was detected at P2. So it
cannot have interacted with the object, and we got information of
the absorbing property of the object {\it without interaction}.
\item[iii.] The particle is detected neither at P1 nor at P2. Here we
conclude,
that the particle was absorbed by a black object, which, clearly,
involved an interaction.

\end{description}

The result in ii is a nice example of the wave-particle-duality.
The wave description determines the probabilities of the detection at
the ports P1 and P2 by interference of the wavefunctions of the two
paths. The particle description lets us conclude, that the particle
{\it did not interact} with the black object when it was detected at P2:
If it
had been at the site of the black object, it would have been caught
there, and could not have reached P2.

The three outcomes permit putting the tested objects into three groups
as sketched in Fig.2. Groups ii and iii will only contain black
objects. All transparent
objects are accumulated in group i, but also some black objects, because
in a single test the
probability for a transparent object to be placed there is 100\%, while
for a black object it is 25\%. With repeated tests of the objects
of this group, the probability of still having a black
object there will approach zero. Simultaneously, the
number of objects in group ii will reach one third of the originally
available black objects. Two thirds of the black objects will
end up in group iii, because they interacted with the test particle and
absorbed it.

The purpose of the present experiment was to perform such a selection
procedure with a real neutron interferometer. The black objects were
realized as a neutron detector. In this manner the three possible
outcomes of an individual test could be observed. For transparent
objects the interferometric paths were left empty.

\section{Experiment and Results}

The experiment was performed with the perfect-silicon-crystal neutron
interferometer installed at the north beam port of the 250kW TRIGA MARK
II reactor of the Atominstitut. This interferometer separates the
paths by about 4 cm and recombines them again. The action of
semitransparent
mirrors is accomplished through Bragg-diffraction in Laue geometry at
the 220-planes of the four crystal slabs. As can be seen in Fig.3, the
basic layout is similar to a Mach-Zehnder interferometer and therefore
well suited to test the EV-scheme. The main difference to
the ideal case is that the two output beams are not equivalent. Output
beam {\bf 1} can have a contrast of 100\% (i.e. be fully modulated), but
output beam {\bf 2} cannot \cite{Petrascheck}, because the number of
reflections and transmissions at the crystal slabs is not the same for
the two paths making up output beam {\bf 2}. But due to crystal
imperfections, even output beam {\bf 1} never reaches a contrast of
100\%. A contrast of 40\% is obtained with a cross section of 4 mm
(width) times 7 mm (height) of the incident beam, and a wavelength of
$\lambda = 1.8 \AA$ with a spread of $\Delta \lambda / \lambda \approx
1\%$. When normalizing
the average intensity at output {\bf 1} to 1, the intensities of our set
up are, as a function of a phase shift $\phi$ between paths {\bf I} and
{\bf II} (setup as shown in Fig.3, but without the Cd sheet and without
an object in path {\bf I}):
\be
I_1(\phi) = 1 + 0.4 \cos(\phi)
\ee
and
\be
I_2(\phi) = 1.7 - 0.4 \cos(\phi).
\ee
There is no phase shift $\phi$ at which one of the output beams is
completely dark. This is contrary to the crucial assumption
in the EV scheme. Consequently, a statement of the sort "if a neutron is
detected at P2, a black object is with certainty in one of
the arms of the intereferometer" was no longer possible. Nevertheless
each test had to lead to a decision into which group the tested
object should be put. We kept the three groups outlined above. But
the certain identification of black objects had to be replaced by a
probabilistic one, and it was attempted to achieve an enrichment of
transparent objects in one group and of black objects in the two other
groups. For this purpose theoretical simulations were undertaken, using
the performance characteristics of the actual interferometer
\cite{Diplom.Meinrad}. The questions to be answered were, whether the
probability of successful interaction-free identification of black
objects could be increased by

\begin{description}
\item[(a)] attenuating the intensity of beam {\bf I}, into which the
test objects were to be placed, with an additional partial absorber, and
\item[(b)] introducing an additional constant phase shift of $\pi$, such
that a registration of a neutron at P1 rather than at P2 could
be taken to indicate the presence of a black object.
\end{description}

It turned out that the optimal conditions could be obtained when
attenuating beam {\bf I} to $t=16\%$ of its original intensity. This
would
reduce the amplitude of the intensity oscillations at P1 and P2 as a
function of $\phi$ by a factor $\sqrt{t}=0.40$. The attenuation was done
with a Cd-sheet of 125$\mu m$ thickness, which was placed in front of
the second crystal slab, and parallel to this slab, as shown in Fig.3.
The actual transmittance of this sheet was 0.158(4). The simulation also
showed, that the phase shift between beams {\bf I} and {\bf II} should
remain
zero, just as in the EV-scheme, such that the count rate at P1 is at
the {\it maximum} when a transparent object, or none, is present at the
test position. But this, still, made it necessary to set the aluminum
phase shifter to a certain value, because contrary to the ideal case,
the actual interferometer has an internal phase offset due to
inaccuracies of slab thicknesses and distances. In the experiment
the registration of a neutron at P1 was interpreted as the presence of a
transparent object, while registration at P2 was taken to indicate a
black object.

The probabilities of the possible results were determined by measuring
the intensities of the neutron beams at the detectors P1 and P2, and,
when applicable, at the detector D, which constituted the black object.
The measured intensities of a typical run are listed in Table 1 below.
Averaging over several runs was not done, because they showed different
contrast, depending on the vibrational level of the building.
A run lasted for approximately four hours, during which measurements
with and without the
black object in path {\bf I}, as well as background measurements and
normal
interference scan measurements had to be taken \cite{Meinrad.p61}.
Without any object in the paths, but with the attenuating Cd-sheet in
place (Fig.3), the total intensity of P1 {\it plus} P2, including
background, was 1.25 counts/sec.

\vspace{5mm}

\noindent
{\bf Table 1:} Observed neutron counts with transparent (=none) or
black object in the test position.

\vspace{3mm}

\begin{tabular}{|c||r|r||r|} \hline
detector & transparent object & black object & background \\  \hline
\hline
P1  &   3561  &   2073 &  215  \\ \hline
P2  &    793  &    999 &   59  \\ \hline
D   &    ---  &   2253 & 1422  \\ \hline
\end{tabular}

\vspace{5mm}

The background was determined by rotating the interferometer crystal
around the vertical axis away from the Bragg condition by some 3
degrees, such that the incident neutron beam went straight through the
first and second crystal slabs. The background at D is due to its
position close to the incident beam, whose intensity is three orders of
magnitude larger than the intensity which the Bragg condition selects
for the interferometer. The background at D is in part also due to the
gamma radiation which is created when neutrons are absorbed in the Cd
sheet in a nuclear reaction.
The detectors P1 and P2 were standard $BF_3$-filled (atmospheric
pressure) cylindrical neutron detectors with a diameter of 5 cm and a
length of around 40 cm, and were hit axially by the respective beams.
Their efficiencies exceeded 97\%. The detector D, representing the black %%@
object, had
an efficiency of only 65\%. Its outside diameter was 2.54 cm, and its
length some 30 cm. When at the test position, beam {\bf I} hit
it perpendicular to the axis of the cylinder.
 It was filled with four atmospheres of $He^3$. The
capture cross section of a thermal neutron at a $He^3$ nucleus is 5330
barns. This results in a probability of 0.76 that a neutron will be
captured in the gas.
The reduction to 0.65 is due to averaging over the beam cross
section and to electronic discrimination, which was needed to suppress
as much as possible the unwanted gamma-background.
If the neutron went through detector D unaffected, it was certainly
absorbed by the shielding material behind the detector. Thereby it was %%@
ensured, that the black object really was black.

From the data in table 1 it is possible to infer whether an actual
interaction-free identification of black objects is possible. In so
doing we will assume 100\% efficiency of detectors P1 and P2. Since this
is not very different from their actual efficiency the resulting error
will be small. Detector D of the black object is not needed for the
analysis. This detector only served for a check of the
consistency of the results.

After subtracting the background counts, which is permissible,
because they can in principle be made arbitrarily low with improved
shielding and electronics, the probabilities for an object to be put
into one of the three groups after a test with only a single
particle, can be calculated as given in table 2:

\vspace{5mm}

\noindent
{\bf Table 2:} Probabilities of detection of neutron at one of the
detectors with black or transparent object in test position.
Standard deviations include only Poisson statistics of the counts of
table 1. Numbers in
rectangular brackets are probabilities for ideal Mach-Zehnder
interferometer.

\vspace{3mm}

\begin{tabular}{|c||c|c|c|} \hline
object &  detection at P1 (i) &  detection at P2 (ii) & absorption in
object (iii) \\ \hline \hline
black        &    .455$\pm$.014 [.25] &  .231$\pm$.009 [.25] &
.314$pm$.018 [.50] \\ \hline
transparent  &    .820$\pm$.020 [1]   &  .180$\pm$.008 [0]            &
--- \\ \hline \end{tabular}

\vspace{5mm}

It is noteworthy, that the probability that the neutron gets absorbed in
the black object is significantly lower than in an ideal Mach-Zehnder
interferometer. This improved performance is also retained in the limit
of infinitely many tests of the objects
remaining in group i. In such a procedure, the probability for a black
object {\it not to
interact} with the neutron and ultimately to be put into group ii is given %%@
by \be
p_{black}^{(ii)} \sum_{n=0}^{\infty}\left[ p_{black}^{(i)}\right]^n \approx %%@
.424 ,
\ee
where we have inserted from table 2, $p_{black}^{(i)} = .455$ and %%@
$p_{black}^{(ii)} = .231$. Correspondingly, the probability that the black %%@
object ultimately absorbs a particle in this procedure and thus is put into %%@
group iii is $1-.424=.576$. But, unfortunately, in such
repeated tests the transparent
objects, too, will accumulate in group ii. The reason is, that their %%@
probability of being put into group i in a single test is less than 1, %%@
namely $p_{trans}^{(i)}=.820$, according to table 2.
Therefore, when testing objects in group i again and again, until either P1 %%@
or D fires, all transparent objects and 42.4\% of the black objects will end %%@
up in group ii.

A closer analysis reveals, that with our real neutron interferometer, the
best separation of black and transparent objects is obtained when
testing each object {\it only once}. Then one can obtain a separation
as shown in Fig.4. The x-axis represents the fraction of black objects
in the original ensemble. The full curves going from the lower left to
the upper right corner represent the fraction of black objects ending
up in group ii, which is the group where
the tested object is put when the particle is detected at P2. The
thick line is the most likely value, the two
thin ones delimit its standard deviation. The dashed line indicates
"no separation" of black and transparent objects. One notes
that an enrichment of black objects in group ii does happen, because
their fraction there is always larger than their fraction in the
original ensemble. For instance, for a fraction of .5 of black
objects in the original ensemble, their fraction in the
final ensemble ii will be .562$\pm$.014.
Therefore, even with a neutron interferometer, whose
performance is far from an ideal Mach-Zehnder interferometer, and in
fact far from the best available neutron interferometers, {\it some
information} is obtained in an interaction-free manner.
For the sake of completeness Fig.4 also shows the fraction of
transparent objects in group i after a single test. These are
the curves extending from the upper left to the lower right
corner. As expected, transparent objects are accumulated in group i.

It is also interesting to ask whether for the {\it real} neutron
interferometer there exists a method, different from the EV-method,
which permits arbitrarily good separation of transparent and black
objects. This is indeed possible, albeit at the cost of only a tiny
fraction of the originally available black objects remaining. The method
consists in repeated tests of the objects in {\it group ii}.
As before, let $p_{black}^{(ii)}$ and $p_{trans}^{(ii)}$ denote the
probabilities that a black, respectively transparent, object will be put
into group ii after a single test. According to table 2 we have,
again neglecting experimental uncertainties, $p_{black}^{(ii)} = .231$ and
$p_{trans}^{(ii)} = .180$. When testing objects of group ii a further $N-1$
times, a black object of the original ensemble is therefore $\left(
p_{black}^{(ii)} / p_{trans}^{(ii)} \right)^N = (1.283)^N$ times more likely %%@
to remain in group ii
as compared to a transparent object of the original ensemble.
Arbitrarily good purification of group ii is therefore possible {\it without
interaction.} The number of black objects group ii contains in the end, will %%@
however only
be $\left( p_{black}^{(ii)} \right)^N$ times the number of black objects in %%@
the original ensemble.

\section{Conclusion}

We have performed a test of interaction-free measurement with a neutron
interferometer of the Mach-Zehnder type. In the unobstructed
interferometer the probability of a neutron to be found in the path of
the test position for the black and the transparent objects to be
identified was around .3, and correspondingly the probability to find
it in the other path was around .7. The visibility contrast at the exit
port P1 only reached around 40\%. With such strong deviations from the
characteristics of an ideal Mach-Zehnder interferometer it was
nevertheless possible to show that an original ensemble with unknown
proportions of black and transparent objects can be separated into two
groups by testing each object with essentially only one neutron (about
1.04 neutrons on average), which must not be absorbed in the test object. %%@
Then
one of these groups is guaranteed to contain a higher proportion of
black objects than the original ensemble. The black objects are laid
out as a neutron detector plus absorptive shielding. A neutron interacting
with a black object would certainly be absorbed by it. Therefore, the
result shows interaction-free measurement at work.

\section{Acknowledgment}

We would like to thank Professor H. Rauch for permission to use his
neutron interferometry setup and to adapt it to the needs of this
experiment.

\pagebreak
{\bf Figure Captions}

\vspace{5mm}

\noindent
{\bf Fig.1: (a)} The test particle will exit the interferometer at P1
when there is no object, or a perfectly transparent one, in
one path of the interferometer.
{\bf (b)} With the perfectly absorbing object D in one path, the test
particle, if it is not absorbed, can exit at P1 as well as at P2.

\vspace{5mm}

\noindent
{\bf Fig.2:} An original ensemble of black and transparent objects can
be separated into three groups by a test in the interferometer. The
black objects in group ii get there without having interacted with the
test particle.

\vspace{5mm}

\noindent
{\bf Fig.3:} Experimental scheme. The base area of the four plate
interferometer is 144mm $\times$ 100mm. Crystal slabs, phase shifter,
black object (= detector D plus shielding) and beam widths are drawn to
scale. The wide incident beam is collimated by a boron carbide (BC)
slit. The transmitted part of beam I after the second crystal slab, and
that of beam II after the third crystal slab, leave the interferometer
unused. Detectors at exit beams P1 and P2 not to scale.

\vspace{5mm}

\noindent
{\bf Fig.4:} Results of experiment. Full lines (mean, and plus minus
one standard deviation): In a single test of each object, a
black object is more likely to be put into group ii, unless the neutron
is absorbed by it in an interaction, while a transparent object is more
likely to be put into group i. Dashed lines: What could be achieved when
randomly putting objects of the original ensemble into different groups.


\begin{thebibliography}{99}

\bibitem{Dicke} R. H. Dicke, Am. J. Phys. {\bf 49}, 925 (1981).
\bibitem{EV1} A. C. Elitzur and L. Vaidman, Found. Phys. {\bf 23}, 987
(1993).
\bibitem{EV2} L. Vaidman, Quantum Opt. {\bf 6}, 119 (1984).
\bibitem{Bennet} C. H. Bennet, Nature {\bf 371}, 479 (1994).
\bibitem{Vaidman.quant-ph} L. Vaidman,
http://xxx.lanl.gov/abs/quant-ph/9610033.
\bibitem{Zeilinger.idea} P. Kwiat, H. Weinfurter, T. Herzog, A.
Zeilinger, and M. Kasevich, Ann. N.Y. Acad. Sci. {\bf 755}, 383 (1995).
\bibitem{Paul.Pavicic} H. Paul and M. Pavicic, J. Opt. Soc. Am. {\bf B
14}, 1275 (1997); M. Pavicic, Phys. Lett. {\bf A 223}, 241 (1996).
\bibitem{Zeilinger.exp} P. Kwiat, H. Weinfurter, T. Herzog,
A. Zeilinger, M. Kasevich, Phys. Rev. Lett. {\bf 74}, 4763 (1995).
\bibitem{Kwiat.review} P. Kwiat, H. Weinfurter and A.
Zeilinger, Scientific American, Nov. 1996, p. 52.
\bibitem{Krenn} G. Krenn, J. Summhammer, and K. Svozil, Phys. Rev. {\bf %%@
A53}, 1228 (1996).
\bibitem{Karlsson} A. Karlsson, G. Bj\"ork, E. Forsberg,
http://xxx.lanl.gov/abs/quant-ph/9705006.
\bibitem{Petrascheck} D. Petrascheck, Phys. Rev. {\bf B 35}, 6549
(1987), and references therein.
\bibitem{Diplom.Meinrad} M. Hafner, Diploma thesis,
Faculty of Natural Sciences, Technical University of Vienna, 1996 (in
German; unpublished).
\bibitem{Meinrad.p61} ibid., p.61, ff.
\end {thebibliography}
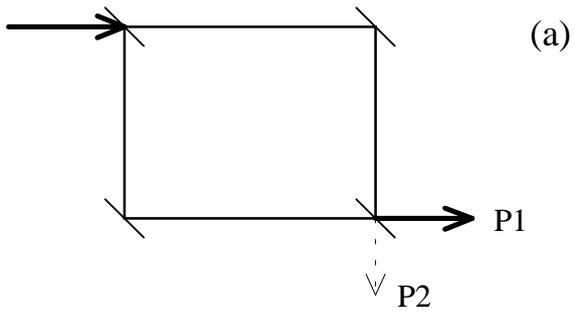
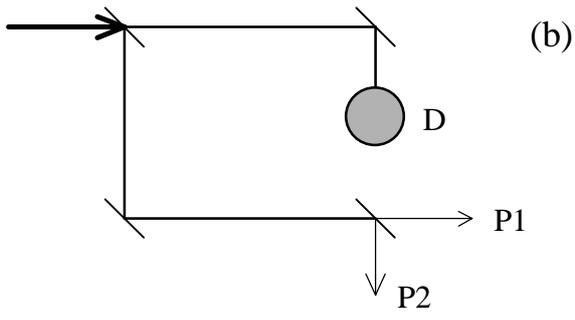
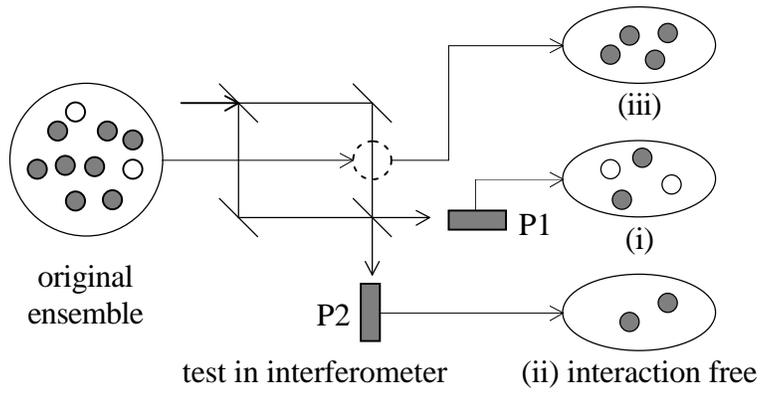
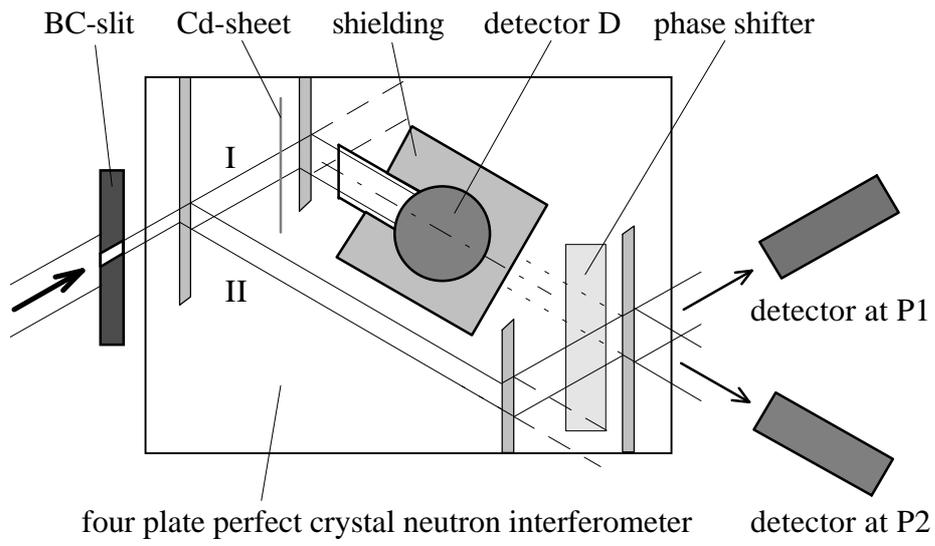
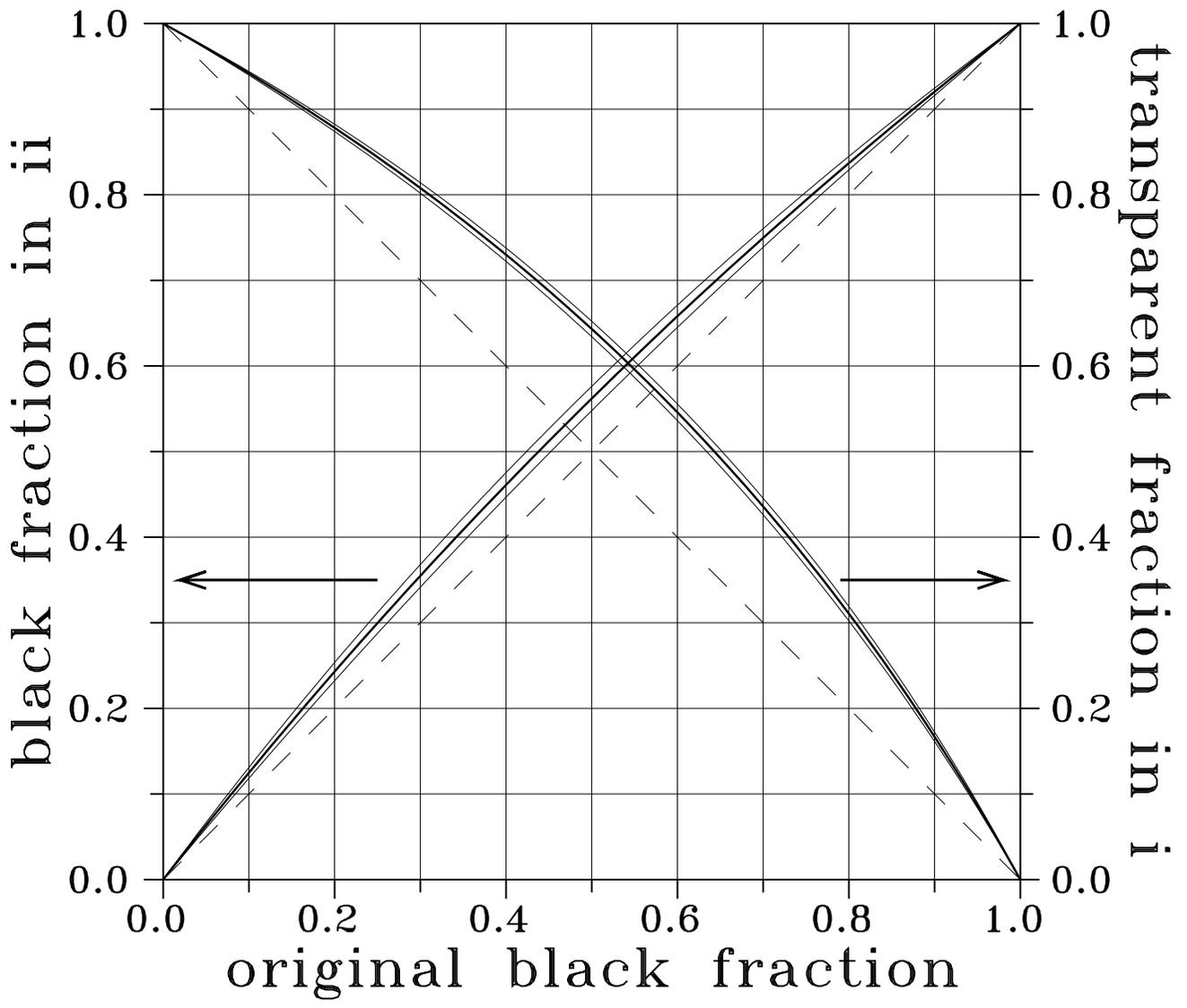
\end{document}